\documentstyle[12pt]{article}
\begin{document}
\def\bib#1{[{\ref{#1}}]}
\def\at{\tilde{a}}

\begin{titlepage}
         \title{The Generalized Uncertainty Principle  \\
                from Quantum Geometry}
\author{S. Capozziello$^{a}$\thanks{E-mail:
capozziello@physics.unisa.it}, ~~G. Lambiase$^{a}$\thanks{E-mail:
lambiase@physics.unisa.it}, ~~and G. Scarpetta$^{a,b}$\thanks{E-mail:
scarpetta@physics.unisa.it} \\
 {\em $^a$Dipartimento di Scienze Fisiche ``E.R.
 Caianiello''}\\ {\em  Universit\`a di Salerno, 84081 Baronissi (SA), Italy} \\
 {\em $^a$Istituto Nazionale di Fisica Nucleare, Sezione di Napoli} \\
 {\em $^b$International Institute for Advanced Scientific Studies} \\
 {\em Vietri sul Mare (SA), Italy.} }

             \date{\empty}
              \maketitle

              \begin{abstract}
The generalized uncertainty principle of string theory is derived in the framework of
Quantum Geometry by taking into account the existence of an upper limit on the
acceleration of massive particles.
          \end{abstract}

\thispagestyle{empty}
\vspace{20. mm}
              \vfill
          \end{titlepage}

\section{Introduction}

The problem of reconciling Quantum Mechanics (QM) with General Relativity is one of the
task of modern theoretical physics which, until now, has not yet found a consistent and
satisfactory solution. The difficulty arises since general relativity deals with the
events which define the world--lines of particles, while QM does not allow the
definition of trajectory; in fact the determination of the position of a quantum
particle involves a measurement which introduces an uncertainty into its momentum
(Wigner, 1957; Saleker, 1958; Feynman, 1965).

These conceptual difficulties have their origin, as argued in Ref. (Candelas, 1983;
Donoghue, 1984, 1985), in the violation, at quantum level, of the weak principle of
equivalence on which general relativity is based. Such a problem becomes more involved
in the formulation of quantum theory of gravity, owing to the non--renormalizability of
general relativity when one quantizes it as a local Quantum Field Theory (QFT) (Birrel,
1982) .

Nevertheless, one of the most interesting consequences
of this unification is
that in quantum gravity there exists a minimal observable distance on the order
of the Planck distance, $l_P=\sqrt{G\hbar/c^3}\sim 10^{-33}$cm, where $G$ is
the Newton constant. The existence of such a fundamental length is a
dynamical phenomenon due to the fact that, at Planck scales, there are {\it
fluctuations} of the background metric, {\it i.e.} a limit of the order of
Planck length appears when quantum fluctuations of the gravitational field are
taken into account.

In absence of a theory of quantum gravity, one tries to analyze quantum aspects of
gravity retaining the gravitational field as a classical background, described by
general relativity, and interacting with matter field. This {\it semiclassical
approximation} leads to QFT and QM in curved space-time and may be considered as a
preliminary step towards the complete quantum theory of gravity. In other words, we take
into account a theory where geometry is classically defined while the source of Einstein
equations is an effective stress--energy tensor where contributions of matter quantum
fields, gravity self--interactions, and quantum matter--gravity interactions appear
(Birrel, 1982).

Besides, the canonical commutation relations between the momentum
operator $p^{\nu}$ and position operator $x^{\mu}$, which in Minkowski
space-time are
$[x^{\mu}, p^{\nu}]=i\hbar\eta^{\mu\nu}$, in a curved space-time
with metric $g_{\mu\nu}$ can be generalized as
\begin{equation}\label{1.1}
[x^{\mu}, p^{\nu}]=i\hbar g^{\mu\nu}(x)\,{.}
\end{equation}
Eq. (\ref{1.1}) contains gravitational effects of a particle in first quantization
scheme. Its validity is confined to curved space--time asymptotically flat so that the
tensor metric can be decomposed  as $g_{\mu\nu}=\eta_{\mu\nu} +h_{\mu\nu}$, where
$h_{\mu\nu}$ is the (local) perturbation to the flat background (Ashtekar, 1990). We
note that the usual commutation relations between position and momentum operators in
Minkowsky space--time are obtained by using the veirbein formalism, i.e. by projecting
the commutator and the metric tensor on the tangent space.

As it is well known, a theory containing a fundamental length on the order of $l_P$
(which can be related to the extension of particles) is string theory. It provides a
consistent theory of quantum gravity and allows to avoid the above mentioned
difficulties. In fact, unlike point particle theories, the existence of a fundamental
length plays the role of natural cut--off. In such a way the, ultraviolet divergencies
are avoided without appealing to the renormalization and regularization schemes (Green,
1987).

Besides, by studying string collisions at planckian energies and through a
renormalization group type analysis (Veneziano, 199; Amati, 1987, 1988, 1989, 1990;
Gross, 1987, 1988; Konishi, 1990; Guida, 1991; Yonega, 1989), the emergence of a minimal
observable distance yields to the generalized uncertainty principle
\begin{equation}\label{1.2}
\Delta x\geq\frac{\hbar}{2\Delta p}+\frac{\alpha}{c^3}G\Delta p\,{.}
\end{equation}
Here, $\alpha$ is a constant. At energy much below the Planck
mass, $m_P=\sqrt{\hbar c/G}\sim 10^{19}$GeV/c$^2$, the extra term in Eq.
(\ref{1.2}) is irrelevant and the Heisenberg relation is recovered, while, as we
approach the Plack energy, this term becomes relevant and, as said, it is
related to the minimal observable length.

The purpose of this paper is to recover the generalized uncertainty principle, Eq.
(\ref{1.2}), in the framework of Quantum Geometry theory (Caianiello, 1979, 1980a,
1980b, 1992). It tries to incorporate quantum aspects into space--time geometry so that
one--particle QM may acquire a geometric interpretation. Its formulation is based on the
fact that the position and momentum operators are represented as covariant derivatives
with an appropriate connection in the eight--dimensional manifold and the quantization
is geometrically interpreted as curvature of phase space.

A consequence of this geometric approach  is the existence of maximal acceleration
defined as the upper limit to the proper acceleration ${\cal A}$ experienced by massive
particles along their worldlines (Caianiello, 1981, 1982, 1984). It can be interpreted
as mass--dependent, ${\cal A}_m=2mc^3/\hbar$ ($m$ is the mass of particle), or as an
universal constant, ${\cal A}=m_Pc^3/\hbar$ ($m_P$ is the Planck mass). Since the regime
of validity of (\ref{1.2}) is at Planck scales, in order to derive it from quantum
geometry, we will consider maximal acceleration depending on Planck mass.

The existence of a maximal acceleration has several implications for relativistic
kinematics (Scarpetta, 1984), energy spectrum of a uniformly accelerated particle
(Caianiello, 1990a), Schwarzschild horizon (Gasperini, 1989), expansion of the very
early universe (Caianiello, 1991), tunneling from {\it nothing} (Capozziello, 1993;
Caianiello, 1994), and mass of the Higgs boson (Kuwata, 1996). It also makes the metric
observer--dependent, as conjectured by Gibbons and Hawking (Gibbons, 1977) and leads, in
a natural way, to the hadronic confinement (Caianiello, 1988). Besides, the regularizing
properties of the maximal acceleration has been recently analyzed in Ref. (Feoli, 1999),
and its applications in the framework of string theory have been studied in Refs.
(Feoli, 1993; McGuigan, 1994).

Moreover, concrete experimental tests of the consequence of the maximal acceleration
have been proposed in Refs. (Caianiello, 1990; Papini, 1995a; Lambiase, 1998).

Limiting values for the acceleration were also derived by several authors on different
grounds and applied to many branches of physics (Brandt, 1983, 1984, 1989; Das, 1980;
Frolov, 1991; Papini, 1992, 1995b; Pati, 1992; Sanchez, 1993, Toller, 1988, 1990, 1991;
Vigier, 1991, Wood, 1989, 1992).

The paper is organized as follows.
In Section 2, we shortly discuss quantum geometry theory,
recalling only the main topics used in this paper.
Section 3 is devoted to
derive the generalized uncertainty principle from quantum geometry.
Conclusions are discussed in Section 4.

\section{\bf Quantum Geometry Theory}

Quantum geometry includes the effects of
the maximal acceleration on dynamics of particles in enlarging
the space-time manifold
to an eight-dimensional space-time tangent bundle TM, {\it i.e.}
$M_8=V_4\otimes TV_4$, where $V_4$ is the background space--time equipped with
metric $g_{\mu\nu}$.
In this way, the invariant line element defined in $M_8$ is generalized as
\begin{equation}\label{2.1}
d\tilde{s}^{2}=g_{AB}dX^{A}dX^{B},\quad A,B=1,\ldots,8\,{,}
\end{equation}
where the coordinates of $M_8$ are
\begin{equation}\label{2.2}
X^{A}=\left(x^{\mu};{\frac{c^2}{\cal A}}{\frac{dx^{\mu}}{ds}}\right),\quad
\mu=1,\ldots,4\,{.}
\end{equation}
$ds$ is the usual infinitesimal element line, $ds^2=g_{\mu\nu}dx^{\mu}
dx^{\nu}$, ${\cal A}$ is the maximal acceleration and
\begin{equation}\label{2.3}
g_{AB}=g_{\mu\nu}\otimes g_{\mu\nu}\,{.}
\end{equation}
From Eq. (\ref{2.3}), it follows that the generalized line element
(\ref{2.1}) can be written as
\begin{equation}\label{2.4}
d\tilde{s}^{2}=g_{\mu\nu}(dx^{\mu}dx^{\nu}+\frac{c^4}{{\cal A}^2}d\dot{x}^{\mu}
d\dot{x}^{\nu})\,{.}
\end{equation}
An embedding procedure can be developed (Caianiello, 1990b) in order to find the
effective space-time geometry where a particle moves when the constraint of the maximal
acceleration is present. In fact, if we find the parametric equations that relate the
velocity field $\dot{x}^{\mu}$ to the first four coordinates $x^{\mu}$, we can calculate
the effective four dimensional metric $\tilde {g}_{\mu\nu}$ induced on the hypersurface
locally embedded in $M_8$. For a particle of mass $m$ accelerating along its worldline,
Eq. (\ref{2.4}) implies that it behaves dynamically as if it is embedded in a
space--time with the metric
\begin{equation}\label{2.5}
d\tilde{s}^2=\left(1+c^4\frac{\ddot{x}^{\sigma}\ddot{x}_{\sigma}}
{{\cal A}^2}\right)ds^2\,{,}
\end{equation}
or, in terms of metric tensor
\begin{equation}\label{2.6}
\tilde{g}_{\mu\nu}=\left(1+c^4\frac{\ddot{x}^{\sigma}\ddot{x}_{\sigma}}
{{\cal A}^2}\right)g_{\mu\nu}\,{,}
\end{equation}
that depends on the squared length of the (spacelike) four--acceleration,
$|\ddot{x}|^2=g_{\sigma\rho}\ddot{x}^{\sigma}\ddot{x}^{\rho}$.
Particularly interesting is
the case in the absence of gravity, $g_{\mu\nu}=\eta_{\mu\nu}$
which corresponds to a flat background. In this case,
any accelerating particle experiences a gravitational field given by
\begin{equation}\label{2.7}
\tilde{g}_{\mu\nu}=\left(1+c^4\frac{\ddot{x}^{\sigma}\ddot{x}_{\sigma}}
{{\cal A}^2}\right)\eta_{\mu\nu}=
\eta_{\mu\nu}+h_{\mu\nu}\,{,}
\end{equation}
where $h_{\mu\nu}=c^4(\ddot{x}^{\sigma}\ddot{x}_{\sigma}/{\cal A}^2)
\eta_{\mu\nu}$ is the quantum (local)
perturbation to the Minkowskian metric. From Eq. (\ref{2.7}) it follows that
\begin{equation}\label{2.8}
\tilde{g}^{\mu\nu}\sim
\left(1-c^4\frac{\ddot{x}^{\sigma}\ddot{x}_{\sigma}}{{\cal A}^2}\right)
\eta^{\mu\nu}\,{.}
\end{equation}

Nevertheless, we stress that this curvature is not induced by matter through
conventional Einstein equation; it is due to the motion in momentum
space and vanishes
in the limit $\hbar\to 0$. Thus, it represents a quantum correction to the given
background geometry, that, henceforth, we will assume flat.

\section{\bf Generalized Uncertainty Principle}

Let us now derive the generalized uncertainty principle (\ref{1.2})
starting from relation (\ref{1.1}), where the tensor metric is
induced by the acceleration of a massive particle in a high energy
scattering process.

According to the hypothesis that microscopic space-time should be regarded as a
four-dimensional hypersurface locally embedded in the larger
height-dimensional manifold,
as discussed in the previous section, accelerated particles
can be associated to four-dimensional
hypersurfaces whose curvature is, in general,
non vanishing. At this semiclassical level, the effective space-time geometry
experimented by interacting particles is curved.

Inserting (\ref{2.7}) into (\ref{1.1}), one gets
\begin{equation}\label{3.1}
[x^{\mu}, p^{\nu}]=i\hbar\left(1+c^4
\frac{(\ddot{x}^{\sigma}\ddot{x}_{\sigma})_m}{{\cal A}^2}
\right)^{-1}\eta^{\mu\nu}\,{.}
\end{equation}
The right-hand side is understood as a {\it c}--function.
The term $(\ddot{x}^{\sigma}\ddot{x}_{\sigma})_m$
is the mean value of the squared length of the four--acceleration which
takes into account the quantum fluctuation of the metric.

Since $\ddot{x}^{\mu}=(1/mc)\delta p^{\mu}/\delta s$,
$\delta p^{\mu}$ is the transferred momentum, it follows that
\begin{equation}\label{3.2}
(\ddot{x}^{\sigma}\ddot{x}_{\sigma})_m\simeq
\frac{1}{m^2c^2\delta s^2}\left[\frac{p^i p^j}{|\vec{p}|^2}
-\delta^{ij}\right](\delta p^i\delta p^j)_m\,{,}\quad i,j=1,2,3\,{,}
\end{equation}
where the high energy limit $E\gg m$ has been used.
Due to the average on the product of transferred momenta,
one can assumes
\begin{equation}\label{3.3}
(\delta p^i \delta p^j)_m\sim \Delta p^2 \delta^{ij}\,{,}
\end{equation}
then Eq. (\ref{3.2}) reads as
\begin{equation}\label{3.4}
(\ddot{x}^{\sigma}\ddot{x}_{\sigma})_m\sim
-2\frac{\Delta p^2}{m^2c^2\delta s^2}\,{.}
\end{equation}
$\Delta p$ is the transferred momentum along the $x$-direction.

As it is well known, two non--commutating operators $A$ and $B$
defined in a Hilbert space, for any given state,
satisfy the uncertainty relation
$$
\Delta A\Delta B\geq \frac{1}{2}\,|<[A, B]>|\,{.}
$$
If $A=x^{\mu}$ and $B=p^{\nu}$, Eqs. (\ref{2.8}) and (\ref{3.1}) allow to
write
\begin{equation}\label{3.5}
\Delta x^{\mu}\Delta p^{\nu}\geq \frac{\hbar}{2}|\eta^{\mu\nu}|
\vert 1-c^4\frac{(\ddot{x}^{\sigma}\ddot{x}_{\sigma})_m}{{\cal A}^2}\vert \,{.}
\end{equation}
From Eq. (\ref{3.4}) and for $\mu=\nu=1$, one obtains
\begin{equation}\label{3.7}
\Delta x\Delta p \geq \frac{\hbar}{2}+
\frac{\hbar c^2}{m^2{\cal A}^2\delta s^2}\Delta p^2\,{.}
\end{equation}
For ${\cal A}=m_Pc^3/\hbar$, where $m_P=(\hbar c/G)^{1/2}$,
and $\delta s\sim \lambda_c\sim \hbar/mc$, with $\lambda_c$ the Compton
length, it becomes
\begin{equation}\label{3.8}
\Delta x\Delta p\geq \frac{\hbar}{2}+\frac{\alpha}{c^3}G\Delta p^2\,{,}
\end{equation}
that is we recover Eq. (\ref{1.1}). $\alpha$ is a free parameter.
Eq. (\ref{3.8}) is the result which we want:
the geometrical interpretation of QM
through a quantization model formulated in a eight--dimensional manifold,
implying the existence of an upper limit on the acceleration of particles,
leads to the generalized principle of string theory.

It is worthwhile to note that, in the last term of (\ref{3.8}), the
dependence on $\hbar$ disappears. So that this term is not related to
quantum fluctuations but, as the uncertainty principle for strings,
it is due to the intrinsic extension of particles.

\section{Conclusions}

Starting from the uncertainty principle of QM written in a space--time,
where the effective geometry is induced by the acceleration of particles
moving along their worldlines, the generalized uncertainty
principle of string theory has been derived.

In this model we have assumed the maximal acceleration as an universal constant
expressed in term of the Planck mass, which value is
${\cal A} \sim 10^{52}m/sec^2$. As expected, it becomes relevant at very
high energy where the emergence of a minimal observable distance occurs.

Unlike the string theory, in which the extension of particles is introduced
{\it ab initio}, in quantum geometry such an extension is takes into
account through the constraint of the maximal acceleration, that is
by modifying the geometry in which moves an accelerating particle.

In this sense, we can state that the geometrical formulation of
QM is an alternative approach in order to study physics of extended objects.

However, we have to note the fact that we have not used any second
quantization scheme or full QFT approach in deriving our generalized uncertainty
principle, nevertheless it is indicative of the fact that quantum geometry is
an alternative scheme leading to physical interesting results.

\end{document}